\documentclass[11pt]{article}
\usepackage[utf8]{inputenc}
\usepackage[T1]{fontenc}
\usepackage[english]{babel}
\usepackage{amsmath,amsthm,amssymb}
\usepackage{mathtext}
\usepackage{mathptmx}
\usepackage{endnotes}
\let\footnote=\endnote
\usepackage{libertine}

\usepackage{geometry}
\usepackage{csquotes}
\usepackage{natbib}
\usepackage[affil-it]{authblk}

\title{Mechanistic Framework of Global Value Chains}
\author{Sourish Dutta}
\affil{Centre for Development Studies \\ Trivandrum, Kerala}
\date{}

\begin{document}

\maketitle

\section{Motivation}

Indeed, the global production (as a system of creating values) is eventually forming like a gigantic and complex network/web of value chains that explains the transitional structures of global trade and development of the global economy. It's truly a new wave of globalisation, and we term it as the global value chains (GVCs), creating the nexus among firms, workers and consumers around the globe. The emergence of this new scenario asks-- how an economy's firms, producers and workers connect in the global economy. And how are they capturing the gains out of it in terms of different dimensions of economic development?  This GVC approach is very crucial for understanding the organisation of the global industries and firms. It requires the statics and dynamics of diverse players involved in this complex global production network. Its broad notion deals with different global issues (including regional value chains also) from the top down to the bottom up, founding a scope for policy analysis \citep{1}. But it is true that, as \cite{2} points out, any single computational framework is not sufficient to quantification this whole range of economic activities. We should adopt an integrative framework for accurate projection of this dynamic multidimensional phenomenon. 

\section{Introduction}
 The purpose of this paper is to propose a mechanistic structure to capture the systemic mechanics of increasing and complex global production network or global value chains\footnote{\cite{3} uses a variety of terms interchangeably throughout his book to refer the phenomenal hyper-globalisation of economic activities, such as "slicing of the value chain", "fragmentation of the production process", "disintegration of production", "delocalization", "vertical specialization", "global production sharing", "unbundling", "offshoring", "outsourcing" and many more.}. Here I use the term "mechanistic structure" to imply a simplified systemic mathematical construction or description that is framed using comprehensive information about the system of global value chains \citep{4}. In this context, one fundamental question will arise naturally that what did happen to the existing theoretical works and their predictions? As \cite{5} rightly indicates that there’s been a lot of theoretical work, a lot of models that people have written down, but it’s really hard to look at the plant level or firm level and see exactly who is doing it and what factors are driving them. Similarly, \cite{6} draws attention towards a particular limitation in the building blocks (consumer preferences, factor endowments, production technologies) in traditional and new trade theory, where the specification of technology treats the mapping between factors of production and final goods as a black box (Input-output system, which transform given input parameters into output parameters). Besides this, they also identify some gaps in the literature. Firstly, most of the literature in this sphere is static\footnote{Some exceptions are the works of \cite{7} and \cite{8}. \cite{7}'s Ricardian model shows that although the rich country always gains from a reduction in offshoring costs in the long run, the poor country may reduce research effort and therefore suffer in the long run. Whereas, \cite{8}'s Ricardian model of North-South trade shows that incompleteness of international contracts leads to the emergence of product cycles}, and thus many fascinating dynamic questions have not been explored. Secondly, many theories of organisations need a general equilibrium implementation to study robust economic predictions. Thirdly, very little work has been devoted to structurally estimate the existing models\footnote{An exception is the work of \cite{9}. They structurally estimate the multi-country global sourcing model.}. And lastly, there is little concern for welfare impact and policy implications of assessed (quantitatively) changes in the international organisation of production.
 
Indeed, most of the theories are of stylised nature (stylised fact-driven or gray box models), which in turn urges us to conceive a best systemic framework (mechanistic or white box model) is required to realise all those gaps in the global value chains literature. The best systemic framework will view this hyper-globalisation phenomenon as a web of relationships among different components that thereby constitute the larger system of value chains throughout the world. Here we need the best one because it should be the simplest framework that still serves its purpose, that is, which is still complex enough to help us understand the system of global value chains and to solve problems. Perceived in terms of a simple framework, the complexity of this complex global value chains will no longer haze our perspective, and we will virtually be able to observe through the complexity of this larger value chains system at the heart of things \citep{4}. 

We can easily realise that we need a conceptual framework (having economical reality) to build the purpose-oriented mechanistic framework for the complex global value chains. We can get this conceptual framework from USAID (U.S. Agency for International Development). Because USAID applies the broader value chain perspective with the framework of inclusive market systems development to drive economic prosperity through the integration of large numbers of bottom level enterprises into increasingly competitive and complex value chains. Besides this, we will rely on the systems and control theory to frame the mechanistic structure. 

\section{Conceptual Framework}

The value chain describes the full range of activities and services that firms and workers perform to bring a product from its conception to sale in its end use and beyond (final markets -- whether local, national, regional or global). This includes many economic agents such as inputs suppliers, producers, processors or traders, and buyers with different activities such as design, production, marketing, distribution and support to the final consumer. The activities that comprise a value chain can be contained within a single firm or divided among different firms. In the context of globalisation, the activities that constitute a value chain have generally been carried out in inter-firm and intra-firm networks on a global scale. As a result, these international production networks are obviously highly complex in terms of geography, technology, and the variety of types of firms involved – from large retailers to highly mechanised large-scale manufacturers to small, even home-based, production as well as a variety of intra-firm and inter-firm linkages. Sometimes it may be impossible even to identify all the countries that are involved or the extent of their involvement \citep{1}. 

Value chains are situated within broader economic systems. For a systematic framework that expresses the context for value chains, we have to adopt an unique market systems framework, which is proposed by \cite{10} for the USAID. 

\subsection{Market Systems Framework}
 
 The systemic framework of market should, indeed, be fundamental enough to facilitate inclusive economic prosperity (in terms of competitiveness, inclusiveness, and resiliency)\footnote{\textbf{Competitive} -- system actors are able to effectively innovate, upgrade and add value to their products and services to match market demand and maintain or grow market share; \textbf{Inclusive} -- delivering a sustainable flow of benefits to a range of actors, including the poor and otherwise marginalised, as well as to society as a whole; and \textbf{Resilient} -- system actors are able to address, absorb and overcome shocks in the market, policy environment, resource base or other aspect of the system.} at any level. Therefore, it should follow some meta-economic principles, by studying deeper structural and functional aspects of the market systems, understood as a holistic, dynamic, complex, regulated, evolving, and adaptive system of value chains. In fact, these principles interdepend, overlap, and cross-refer with one another \citep{11}. In other words, market systems should include value chains, households \& communities, interconnect with other systems, soft boundaries to become a self-organising complex structure \citep{10}. 
 
 \begin{itemize}
     \item \textbf{Market systems include value chains} -- The market system, shows how households interact with multiple, possibly interconnected value chains. Value chains catalyse and are impacted by broader economic change through multiplier effects.
     
     \item \textbf{Market systems interconnect with other systems} -- Market systems interact with other systems, such as health systems, education systems, socio-cultural systems, and ecosystems. Changes in one system can affect the functioning of other systems. Though no single intervention can be expected to simultaneously transform multiple systems, such networking sometimes allows us to trigger broad-scale change in the market system by targeting linkages with other systems. 
     
     \item \textbf{Market systems include households and communities} -- Households and communities are also systems. Decisions about resource allocation are negotiated among household and community members, influenced by individuals' incentives and expectations, status and decision-making power, a range of socio-cultural norms and traditions, and physical factors that constrain available options. Understanding household and community systems and how they interact with each other and other systems can be important for achieving development objectives.
     
     \item \textbf{Market systems have "soft" boundaries} -- Since market systems contain sub-systems (household systems, value chain systems, etc.), and are connected to health systems, political systems, ecosystems, and the like, we face the challenge of how to define the intervention space. How wide or focused should the parameters be for the purposes of analysis, the design of interventions, the monitoring of systems change, and the evaluation of impact?
     
     \item \textbf{Market systems are complex} -- Market systems contain many actors, institutions, structures and influences that are both interconnected and independent. These system elements interact in ways that are often unpredictable at the transaction level—although at the aggregate level may be observed to follow patterns over time. Consequently, the results of many developmental interventions in market system cannot be predicted in advance.
     
     \item \textbf{Market systems are self-organising} -- Market systems evolve in response to many external and internal factors and forces. It is extremely important for donors and project implementers attempting to strengthen inclusive market systems to understand the drivers that have resulted in a system’s current way of operating. The vested interests of market actors generally need to be addressed for widespread change to occur. Vested interests may not be apparent at the outset of an intervention, but often emerge as market system changes begin to take hold. Ongoing analysis and learning are therefore essential.
 \end{itemize}

\subsection{Value Chains Framework}

The systemic framework of larger value chains requires to identify the structure of the chain, including all individuals and firms that conduct economic activities by adding value and helping move the product toward the end markets, influences the dynamics of firm behaviour and the dynamics of the value chain, which refers to the determinants of individual and firm behaviour and their effect on the functioning of the chain, influence how well the value chain performs.

\subsubsection{Structural Factors}

The structure of a value chain includes all the firms in the chain and can be characterised in terms of five elements described below:

\begin{itemize}
    \item \textbf{End markets} -- End markets are the focal point of the value chains framework. The term end market is used to specify where the final instance of buying or selling takes place in a value chain. Usually it is where the end-user is located, such that the individual or organisation for whom the product or service has been created, and who is not expected to resell that product or service. For example, creating a consumer product may entail many transactions between various value chain actors, but the end market is where the product becomes available for purchase by the consumer. For a business-related product or service, the end market is where the sale occurs to the organisation that will use the product or service in its own operations. The terms destination market, target market and final market are often used interchangeably with end market. 
    
    The end market determines the characteristics—including price, quality, quantity and timing—of a successful product or service. Depending on its size, an end market can often be divided into smaller market segments according to distinctive characteristics such as price, quality, buyer gender, and geographic location (reflecting regional needs and/or preferences, for example). The potential classifications are numerous and based on a group of people or organisations sharing a combination of characteristics that leads them to exhibit similar purchasing behaviour. If the resulting subgroup is highly differentiated, it may be referred to as a niche market (a portion of a market identified as having a special characteristic that is worth marketing to). It is called market segmentation, and such categorisation of end markets helps sellers focus resources on targeted buyers for their products or services. 
    
    The end market for a product or service can be located anywhere in the world depending on the value chain it flows through. As well, a product or service can flow through multiple value chains reaching many different end markets. For example, a product may simultaneously be sold in a village market by the producer, in major cities of the producer's country by retailers who purchased the product from local traders, in neighbouring countries by regional retailers who purchased the product from an exporter, and in a distant foreign market by retailers who purchased the product from a wholesale importer.
    
    However, local end markets are limited to areas surrounding the source of a product or service. These are typically defined by towns and districts within a country (although in some cases a nearby cross-border market may also be considered a local end market). A national market suggests the product or service is sold throughout a country, or at least in one or more venues beyond local markets. Regional markets encompass an international territory defined by geographic proximity, common language and culture, historic relations, or other characteristic that facilitates trade between countries--examples include Central America, West Africa and the Middle East. And, by definition, global end markets include all potential markets, although in practice often refer to large international markets that are outside of the producing country’s region.
    
    In the systematic value chains framework, it is important to explore each existing and potential end market to determine which offers the greatest benefits (profit margins, embedded services, competitive advantages, etc.) and risks (competition, sizeable investment, fleeting business relationships, etc.) for each actor in the value chain. We have to find a perfect balance between the opportunities and risks of global markets, and the accessibility and limitations of local and regional markets.

    \item \textbf{Business enabling environment} -- System of value chains operate in a business enabling environment (BEE) that can be all at once global, national and local and includes norms and customs, laws, regulations, policies, international trade agreements and public infrastructure (roads, electricity, etc.). The international enabling environment requires the investigation of conventions, treaties, agreements and market standards. While trade agreements can open opportunities for firms, international standards can build obstacles to the same opportunities. Information is also needed at other levels, including national and local policies, duties, business licensing procedures, enacted regulations and the state of public infrastructure. The systemic analysis may need to be further broken down in terms of firm size: there may be particular constraints and opportunities facing micro- and small enterprises (MSEs), for example. Overall, the analysis process must determine whether and how the business enabling environment facilitates or hinders performance of the value chain, and if it hinders, where and how can it be improved. BEE constraints can be difficult to resolve—sometimes requiring considerable time, resources and political capital. 
    
    In addition to these more formal factors, social norms, business culture and local expectations can be powerful aspects of the business enabling environment. Understanding these unwritten rules of society is essential if we want to understand why value chain actors behave the way they do, and reasonably predict how they will behave in response to value chain interventions.
    
    \item \textbf{Vertical linkages} --  Linkages between firms at different levels of the value chain are critical for moving a product or service to the end market. Vertical cooperation reflects the quality of relationships among vertically linked firms up and down the value chain. More efficient transactions among firms that are vertically related in a value chain increase the competitiveness of the entire industry. In addition, vertical linkages facilitate the delivery of benefits and embedded services and the transference of skills and information between firms up and down the chain. MSEs are vertically linked to a varied range of market actors including wholesalers, retailers, exporters, traders, middlemen, input dealers, suppliers, service providers and others. The nature of vertical linkages—including the volume and quality of information and services disseminated—often defines and determines the benefit distribution along the chain and creates incentives for, or constrains, upgrading, defined as “innovation to increase value added.” Moreover, the efficiency of the transactions between vertically linked firms in a value chain affects the competitiveness of the entire industry. An important part of systemic value chain analysis is the identification of weak or missing vertical linkages.
    
    Effective vertical linkages are generally characterised by:
    \begin{itemize}
    \item \textbf{Mutually beneficial relationships}. Symbiotic relationships that benefit all of the actors in a value chain are a major trait of effective vertical linkages. In such a scenario various market actors focus on their own core competencies and through collaborative action realise synergies that improve the competitiveness of the entire chain. Trust long-term joint vision and mutual respect usually form the foundations for developing such relationships.
        
    \item \textbf{Knowledge transfer}. Upgrading of production processes technology equipment management systems etc. is critical for the survival and growth of firms in a competitive marketplace. It is often difficult for small firms to access information about global best practices. Effective vertical linkages facilitate the transfer of knowledge between firms and create the incentives and knowledge platforms required for effective upgrading of MSEs. Prompt information transfers and transparency between vertically linked firms help a value chain respond effectively to changes in market demand.
        
    \item \textbf{Quality standards}. Well-defined widely understood and constantly upgraded quality standards are another defining element of effective vertical linkages. Vertically linked firms are proactive not reactive: Large firms empower and help small firms to understand and adopt the quality standards to meet market demand.
        
    \item \textbf{Embedded services}. The frequent provision of high-quality embedded services (where a service is provided as part of the transaction at no extra cost) typifies effective vertical linkages. Lead firms can provide a wide range of embedded services to affiliated suppliers and buyers to ensure consistent quality of end products and services. These embedded services are often seen as an integral part of business transactions and considered a necessary cost of doing business.
        
    \item \textbf{Financial flows}. Effective vertical linkages are often accompanied by a high volume and variety of financial flows. Larger firms may employ a variety of financial instruments (supplier credit working capital loan leasing services etc.) to support the operations of their linked suppliers.
    \end{itemize}
    
    Nonetheless, the nature of the vertical relationship between buyers and sellers is typically varied and dynamic and affected by end market requirements the business enabling environment product attributes technology socio-economic conditions and competitive pressures.

    \item \textbf{Horizontal linkages} -- Horizontal linkages, both formal as well as informal, between firms at all levels in a value chain can reduce transaction costs, create economies of scale, and contribute to the increased efficiency and competitiveness of an industry. In addition to lowering the cost of inputs and services, horizontal linkages can contribute to shared skills and resources and enhance product quality through common production standards. Such linkages also facilitate collective learning and risk sharing, while increasing the potential for upgrading and innovation. Value chain analysis also considers competition between firms. While cooperation can help firms achieve economies of scale and overcome common constraints to pursue opportunities, competition can encourage innovation and drives firms to upgrade. The most successful horizontal linkages maintain a balance between these two contrasting, but critical and complementary concepts. One of the objectives of value chain analysis is to identify areas where collaborative bargaining power could reduce the cost or increase the benefits to small firms operating in the chain.
    
    \item \textbf{Supporting markets} -- Supporting markets play an important role in firm upgrading. They include financial services; cross-cutting services such as business consulting, legal advice and telecommunications; and sector-specific services, for example, irrigation equipment or handicraft design services. Not all services can be provided as embedded services by value chain actors, and so vibrant supporting markets are often essential to competitiveness. Service providers may include for-profit firms and individuals as well as publicly funded institutions and agencies. Support markets operate within their own value chain—most service providers themselves need supplies, training and financing in addition to strong vertical and horizontal linkages. Value chain analysis should therefore seek to identify opportunities for improved access to services for target value chain actors in such a way that the support markets will be simultaneously strengthened, rather than undermined. Formal supporting markets are likely to expand as the value chain develops. Therefore, when analysing emerging value chains, or ones predominated by MSEs, particular care should be taken to uncover informal sector service providers, which often go unnoticed.
    
\end{itemize}

\subsubsection{Dynamic Factors}

The firms in an industry create the dynamic elements described below through the choices they make in response to the value chain structure.

\begin{itemize}
    \item \textbf{Value chain governance} -- A distinguishing characteristic of value chain analysis is the emphasis not only on the dynamics of end markets but also on the dynamics and shifts in relationships. Value chain governance refers to the relationships among the buyers, sellers, service providers and regulatory institutions that operate within or influence the range of activities required to bring a product or service from inception to its end use. Governance is about power and the ability to exert control along the chain – at any point in the chain, some firm (or organisation or institution) sets and/or enforces parameters under which others in the chain operate. Understanding how and when lead firms set, monitor and enforce rules and standards can help MSEs and other firms in the chain better integrate and coordinate their activities. Governance is particularly important for the generation, transfer and diffusion of knowledge leading to innovation, which enables firms to improve their performance and sustain competitive advantage. Awareness of the governance structure of a value chain can provide governments, donors and development practitioners with information about how best to provide MSEs with the training and technical assistance needed to upgrade their position in the chain. When conducting systemic analysis of value chains, the type of governance structure that exists must be identified since it will contribute significantly to the selection of interventions to increase competitiveness.
    
    \item \textbf{Inter-firm relationships} -- This refers to the nature and quality of the interactions between stakeholders in a value chain. Effective inter-firm relationships are considered an essential component in creating and maintaining value chain competitiveness. Even when other conditions are favorable—end market demand is strong, quality inputs are affordable, technologies are efficient, supporting markets function well, and so on—ineffective relationships can jeopardize the competitiveness of a value chain and its ability to generate economic growth, employment and incomes. Hence, relationships can be supportive of industry competitiveness that enhances MSE benefits or adversarial to it. Supportive relationships facilitate collaboration; enable the transmission of information, skills and services; and provide incentives for upgrading. They are based on a long-term view of the industry, while adversarial relationships are structured to maximise short-term profits. 
    For larger firms at higher levels of the value chain, both cooperation and competition contribute to upgrading and competitiveness. For example, cooperation among exporters can help them to identify common needs and address those needs, through lobbying, branding or other joint activities. At the same time, healthy competition among exporters can foster innovation and promote upgrading. With each firm vying to offer the best quality, design, reliability or other competitive characteristic, the result will be increased efficiency and competitiveness at the industry level.
    
    \item \textbf{Upgrading} --  In order to respond effectively to market opportunities, firms and industries need to innovate to add value to products or services and to make production and marketing processes more efficient. These activities, known as firm-level upgrading, can provide MSEs with higher returns and a steady, more secure income through the development of knowledge and the ability to respond to changing market conditions. Upgrading at the industry-level focuses on increasing the competitiveness of all activities involved in the production, processing and/or marketing of a product or service and mitigating the constraints that limit value chain performance. Upgrading needs to be a continual process and can leads to national economic growth. In value chain analysis, the objective is to identify opportunities and constraints to firm- and industry-level upgrading; specifically the analysis looks for catalyst firms with the incentives, resources and willingness to promote and facilitate upgrading within the chain.
\end{itemize}

\section{Mechanistic Framework}

The system of value chains $\mathbf{S}_l$ can be represented as a 6-tuple:
\begin{equation}
    \mathbf{S}_l = \mathbf{\{C, N, I, B, K, H\}}_l,\qquad \forall\  l = 0, 1, 2,..., m
\end{equation}
where $l$ is an index related to the level of complexity of the value chains system. $l=0$ indicates the top level of this system. For example, when $\mathbf{C}$ is decomposed and each $c_i$ is treated as its own distinct system of value chains, $S_{l,i}$. In fact, the definition recurses with $l=1$, then $2$, and so on.

$\mathbf{C}$ is a multiset of component subsystems of the larger system of value chains, $\{\{c_1, n_1\}, \{c_2, n_2\},...,$ $\{c_i, n_i\},..., \{c_k, n_k\}\}$, where each $c_i$ is a type of structural component of value chains and may also be a subsystem but at level $l+1$. That is, each $c_i$ may itself be defined by a similar 6-tuple in a recursive fashion. Nevertheless, the recursion can prevented from being infinite (In essence, a decomposition stops at a level $l = L$ when the components in $S_L$ are all defined as atomic for the particular kind of system being analysed. Atomic, here, simply means the simplest component, not needing further decomposition because its functions, etc. are already known). Each $n_i$ in the component tuples is an integer enumeration of the number of components of that particular type (multiplicity). These tuples, $\{ c_i , n_i \}$, are multisets. As written here, their cardinality
is one. However, in very complex systems, we can aggregate any number of variations of components of the same basic type (a category)
into a multiset whose cardinality is the number of distinct variations identified. 

$\mathbf{N}$ is a network description in graph theoretical terms. A graph, $\mathbf{G = \{V , E\}}$, is a tuple containing two sets, $\mathbf{V}$ , the vertices or nodes, which in this case correspond to the components in $\mathbf{C}$, and $\mathbf{E}$, the edges or connections between the
nodes. $\mathbf{N}$ provides a map of what components are connected to what other components at this level of complexity

$\mathbf{I}$ is also a graph in which half of the nodes are not inside the system per se but are either sources or sinks or other entity types in the environment. That is, $\mathbf{I}$ contains elements of the environment that are relevant to the system. The other nodes are component subsystems within the system. $\mathbf{I}$ describes the connections a system has with objects in its environment using the same format as $\mathbf{N}$ (which is strictly internal connections).

$\mathbf{B}$ is a complex object that will vary based on details. Fundamentally, it is a description of the set of boundary conditions that maintain the system identity.

$\mathbf{K}$ is also a complex object. This object contains the descriptions of how the various components interact with one another and with the environment. That is, it contains the state transition rules for all interactions. It is an augmentation to the $\mathbf{N}$ and $\mathbf{I}$ graphs. For instance, for each edge $e_i \in N$ , there is an object $k_i$ containing the parameters associated with $e_i$ . For example, $k_i$ could contain a tuple representing the capacity of a flow, the substance flowing, simply the strength of the connection, etc. A similar representation is provided for the graph in $\mathbf{I}$ where augmentation of the connections between external (environmental) elements and internal components is needed. $\mathbf{K}$ is called the system knowledge. It is both self-knowledge (internal structural and functional knowledge) and other knowledge to the extent it details the external connections. Note that the external objects are not, in this definition, modeled. That is, the system contains no explicit knowledge of the internals of external objects, only how they
behave. However, this can be extended in complex adaptive systems to include knowledge of the environmental objects.

Finally, $\mathbf{H}$ is a super complex object that records the history of the system, or its record of state transitions, especially as it develops or evolves. For example, brains learn from experience, and as such their internal microstructures change over time. This is called memory and the current state of $\mathbf{K}$ is based on all previous states. Some simple systems may have a null $\mathbf{H}$; that is, there is no memory of past states. As just mentioned, on the other hand, brains (and indeed all biological systems) have very rich memories. $\mathbf{H}$ is an augment for $\mathbf{K}$.

Each one of these elements is a description of some aspect of the system
and all are needed to provide a complete description. Many of them, such as
$\mathbf{K}$, can be extremely complex. For example, in the case of the brain mentioned above, $\mathbf{K}$ is actually distributed between the genetic inheritance of the individual, which prescribes the major wiring diagram and the major processing modules, and any culturally determined factors that push the direction of development (learning). This is why $\mathbf{K}$ must also contain augments for $\mathbf{I}$. It is also why $\mathbf{H}$ may become extremely complex.

By having a formal definition such as this, it is now possible to apply a great many mathematical tools to analyse many aspects of the system.

\newpage
\theendnotes

\end{document}